\documentclass[aps, twocolumn, prl, showpacs, amsmath, superscriptaddress]{revtex4}
\usepackage{graphicx}
\usepackage{dcolumn}
\usepackage{bm}
\usepackage{float}
\usepackage{amsmath}
\usepackage{subfigure}
\usepackage{amsfonts}
\usepackage{xcolor}
\usepackage{appendix}
\usepackage[colorlinks,
            linkcolor=red,
            anchorcolor=blue,
            citecolor=green
            ]{hyperref}

\begin{document}

\title{Prediction of Ideal Triple Degenerate Points in HfIrAs and HfIrBi}
\author{Zhi-Gang Song}
\affiliation{Max Planck Institute for Chemical Physics of Solids, 01187 Dresden, Germany}
\author{Claudia Felser}
\affiliation{Max Planck Institute for Chemical Physics of Solids, 01187 Dresden, Germany}
\author{Yan Sun}
\email{Yan.Sun@cpfs.mpg.de}
\affiliation{Max Planck Institute for Chemical Physics of Solids, 01187 Dresden, Germany}

\date{\today }

\begin{abstract}

In this study, two ideal triple degenerate point (TDP) semimetals were predicted, namely, HfIrAs and HfIrBi, that exhibit clean TDPs around the fermi level. At the atomic level, the origin of the TDP was analyzed, and when combined with the low-energy effective 4-band model, it was found that the bulk inversion asymmetry directly determines the formation of the TDP for cubic crystals. The projected surface states and corresponding patterns of the fermi arcs with and without tensile strain were investigated for two different crystal face orientations [(111) and (110)], both of which could be detected using photoemission spectroscopy. In addition, the disappearance of the spin texture, which was studied in strained HgTe bulk states, was also observed in the surface states. The results of this study indicate that the two materials are unique platforms on which emergent phenomena, such as the interplay between the TDP of semimetals and superconductivity, can be explored.

\end{abstract}
\maketitle

\section{Introduction}
In the field of condensed matter physics, since the discovery of topological insulators in the past decade, many topological materials have been identified, and range from topological insulators \cite{konig2007quantum,liu2008quantum,hasan2010colloquium,qi2011topological}to
topological crystal insulators \cite{fu2011topological,tanaka2012experimental}, from Dirac semimetals(DSMs) \cite{Liu864,PhysRevB.85.195320,neupane2014observation} to
Weyl semimetals(WSMs) \cite{burkov2011weyl,wan2011topological,weng2015weyl,lv2015experimental}. Not only have these topological materials provided abundant platforms on which to realize unique particles in the field of high energy physics, they have also significantly advanced our understanding of the solid bands in such materials. For DSMs, such as A$_{3}$Bi, where A = Na, K, Rb \cite{Liu864,PhysRevB.85.195320}, and Cd$_{3}$As$_{2}$ \cite{neupane2014observation}, the four-degenerate Dirac points are protected by the $C_{6v}$ and $C_{4v}$ symmetry.
For WSMs, the pair of Weyl points can be viewed as the splitting of four-degenerate Dirac points in momentum space due to the breaking of the inversion or time-reversal symmetries \cite{burkov2011weyl}. Furthermore, in contrast to DSMs and WSMs that are based on four-degenerate Dirac points and two-degenerate Weyl points, respectively, Bradlyn \cite{bradlyn2016beyond} proposed some new fermions in the spirit of crystal symmetry in which the degenerate band crossing points are 3, 6, and 8. In particular, triple degenerate point (TDP) semimetals have attracted more attention as they can be considered as the states between DSMs and WSMs. In fact, a TDP has been observed in HgTe \cite{PhysRevB.87.045202}
along the $[111]$ direction. Sometime later, InAs$_{0.5}$Sb$_{0.5}$ with a CuPt structure \cite{PhysRevLett.117.076403}, was proposed as a novel topological TDP semimetal. Other classes of materials, such as ZrTe, MoC, WC, and WN, have also been proposed as TDP semimetals \cite{PhysRevX.6.031003} and the status of the MoP structure has been confirmed via experiments \cite{lv2017observation}. All of these TDP semimetals can be divided into two classes, called Type A and Type B, based on topologically different invariants accompanied by either one or four nodal lines \cite{PhysRevX.6.031003}. The strained HgTe and InAs$_{0.5}$Sb$_{0.5}$ are Type B TDP semimetals and MoC is a Type A semimetal.

Recently, the existence of many other TDP semimetals \cite{PhysRevB.93.241202,PhysRevB.94.165201,PhysRevB.95.235158,PhysRevB.95.241116,PhysRevLett.119.136401,PhysRevB.96.241204,PhysRevLett.119.256402} have been predicted, including some ternary half-Heusler materials, such as LuPdBi, LuPtBi, and LaPtBi \cite{PhysRevLett.119.136401}.
Since the $[111]$ high symmetry line($\Gamma$-L, $C_{3}$ axis) in the Brillouin Zone (BZ) is protected by $C_{3v}$ symmetry, this facilitates one two-dimensional representation and two one-dimensional representations, which satisfy the requirements for a TDP that has at least one two-dimensional representation and one one-dimensional representation. These ternary half-Heusler materials can be distinguished based on the number of TDP pairs along the $C_{3}$ axis. In the frame of the 6-band Kane model, two key parameters, namely, $C$ and $\Delta$(i.e., the gap between the $\Gamma_{6}$ and $\Gamma_{8}$ states) determine the phase \cite{PhysRevLett.119.136401}. However, the physical mechanism underlying parameter $C$ is not clear. At the same time, other materials in the  NaCu$_{3}$Te$_{2}$ family \cite{PhysRevLett.119.256402,PhysRevB.96.241204} are considered to be ideal TDP semimetals as the TDPs around the fermi level in these materials do not coexist with other quasiparticle bands. Compared with previous half-Heusler materials, these ideal materials facilitate easy exploration of the intrinsic electron structure and transport properties of TDP semimetals. This raises the question of whether ideal TDP semimetals exist in the half-Heusler regime. Based on our research, the answer to this question is positive.

In this study, two half-Heusler materials were predicted, namely, HfIrAs and HfIrBi, that are ideal TDP semimetals, the origin of the TDP was explored at the atomic level, and the physical effect of the $C$ parameter noted in \cite{PhysRevLett.119.136401} was identified via a 4-band effective Hamiltonian model based on the bulk inversion asymmetry (BIA) or Dresselhaus effects \cite{PhysRev.100.580}, which is an essential step in the realization of TDP semimetals in cubic crystals. The projected surface states and fermi arcs for two different crystal face orientations were also investigated with and without tensile strain as the tensile strain will induce a band gap at the $\Gamma$ point and double the number of TDPs, which will result in differences in the bulk and surface states. Finally, the disappearance of the spin texture around critical points in the surface states was also observed as a bulk phenomenon \cite{PhysRevB.87.045202} and was found to be directly determined by changes in the spin winding number.


\section{Results and Discussion}
\begin{figure}
  \centering
  \includegraphics[width=3.5in,height=5.5in]{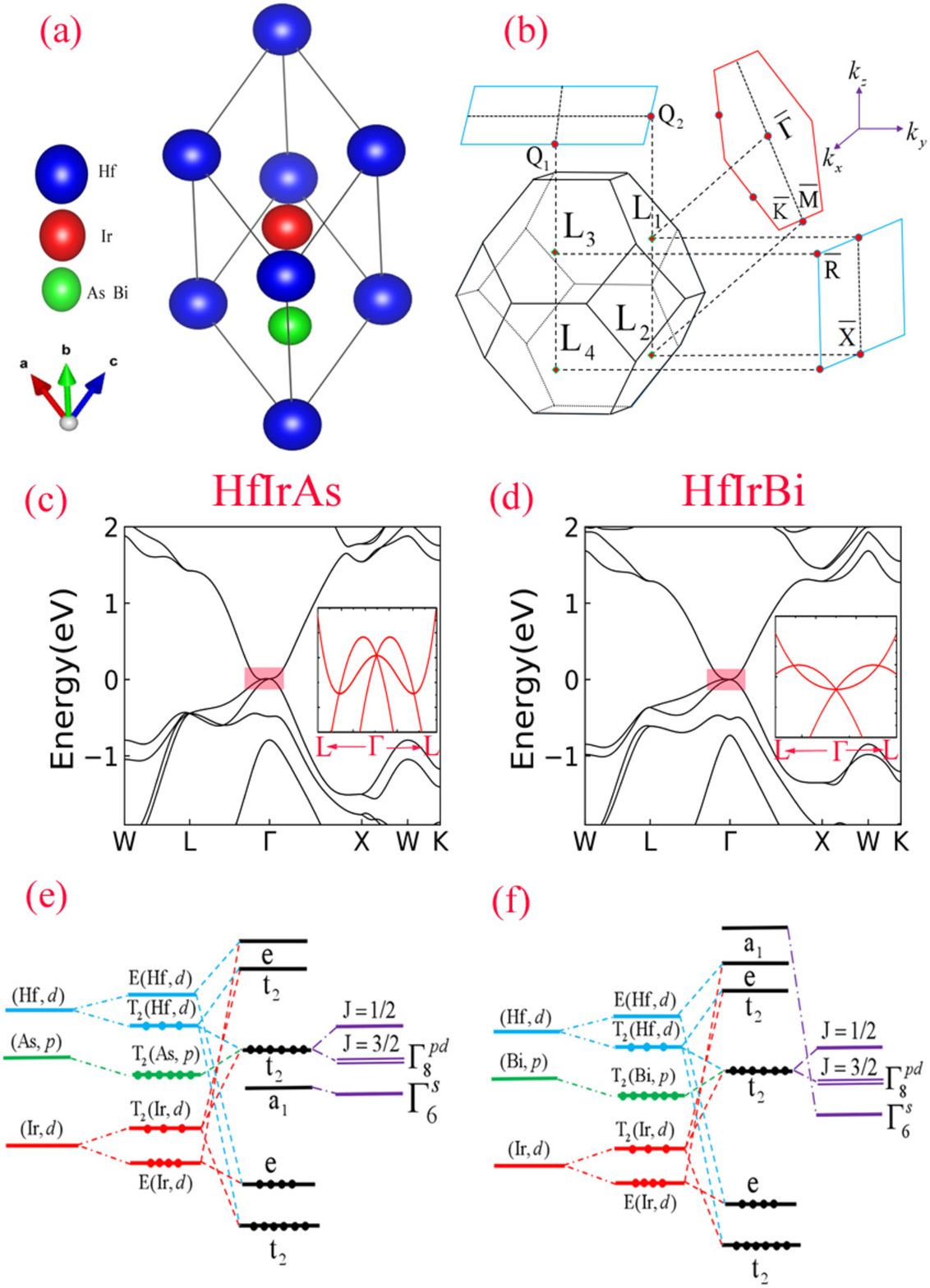}
  \caption{(a)Primitive cell of the HfIrAs and HfIrBi. (b)Schematic of the bulk BZ and surface BZ. The four points L$_{1}$,L$_{2}$,L$_{3}$ and L$_{4}$ are used to label different $C_{3}$ axis($\Gamma$-L lines).
  (c)and (d) show the band structure of HfIrAs and HfIrBi. The insets show enlarged pictures
  of the shallow area. (e) and (f) show schematic diagrams of band evolution in the vicinity
  of $\Gamma$ point for HfIrAs and HfIrBi.
In the cubic structure, each atomic site has $T_{d}$ symmetry.
Hence, the $d$ orbital splits into the T$_{2}$ and E states and the $p$ orbital is in
the T$_{2}$ state. According to the group theory, the hybridization is allowed for the same irreducible representation,
which means that not only the T$_{2}$(E) states, from the nearest neighbours of Hf and Ir strongly couple and repel each
other, but the $p$ states of As(Bi) also contribute as well. The above hybridization and SOC together result in low-ling $\Gamma_{8}^{pd}$ states.
In addition, the hybridized a$_{1}$ state, originating from $s$ states of Hf Ir and As(Bi), can stay above or below low-lying $t_{2}$ state until it splits with the help of SOC. The low-lying branch of splitting is $\Gamma_{6}^{s}$ state(Here the irreducible representations of T$_{2}$ and E are used as Bethe.).}\label{struc_bulk}
\end{figure}

The primitive cell structure of HfIrAs (HfIrBi) is depicted in Fig. \ref{struc_bulk}(a) and the related conventional cell has a noncentrosymmetric face-center-cubic structure with a space $F$-43m(No. 216). This structure has eight $C_{3}$ axes and six mirror planes. The bulk and surface BZs are shown in Fig. \ref{struc_bulk}(b), and the surface BZ is explained later in this paper. The band structure of the HfIrAs and HfIrBi, including the spin-orbital coupling (SOC), can be seen in Figs. \ref{struc_bulk}(c) and \ref{struc_bulk}(d). In the figure, the fermi energy was set to 0 eV. Note that it is only necessary to observe the band structure around the ¦£ point as there is no state near the fermi level in any other areas across the entire BZ. In addition, based on the symmetry analysis, the four bands near the fermi level around the $\Gamma$ point are part of the $\Gamma_{8}$ representation and the lower two bands are part of the $\Gamma_{6}$ representation in the $T_{d}$ point group. To provide a better understanding of the bands,
the band evolution by crystal-field splitting, orbital hybridization, and SOC at the $\Gamma$ point at the atomic level are illustrated in Fig. \ref{struc_bulk}(e) and \ref{struc_bulk}(f). In the cubic $T_{d}$ crystal field, the $d$ states of Hf and Ir will split into three degenerate T$_{2}$ states [T$_{2}$(Hf,$d$) and T$_{2}$(Ir,$d$)] and two degenerate E states [E$_{2}$(Hf,$d$) and E$_{2}$(Ir,$d$)]. Since the transition metal atoms Hf and Ir are mutually tetrahedrally coordinated nearest neighbors, the T$_{2}$(Hf,$d$) and T$_{2}$(Ir,$d$) states, as well as the E$_{2}$(Hf,$d$) and E$_{2}$(Ir,$d$) states, are strongly coupled and repel each other. As a result, there is a large gap between the upper unoccupied $d$ states and the lower occupied $d$ states t$_{2}$.
With the influence of the SOC, the t$_{2}$ states split into four degenerate $\Gamma_{8}^{pd}$ state($\left\vert \frac{3}{2},\pm \frac{3}{2}\right\rangle$ and $\left\vert \frac{3}{2},\pm \frac{1}{2}\right\rangle$) and two-degenerate $\Gamma_{7}^{pd}$ state.
 Meanwhile, the SOC can also further lift the hybridized a$_{1}$ state and induce the $\Gamma_{6}$ state that resides below the lower state. Here, the sites of the $\Gamma_{6}$ state is determined by the strength of the SOC for the case of HfIrBi, as in ScPtBi \cite{chadov2010tunable,gautier2015prediction}, or the lower $s$ state of As in the case of HfIrAs. Apart from the $\Gamma$ point,
 linear and cubic spin splitting can also exist as HgTe due to the absence of inversion symmetry. In terms of the [111] line with the $C_{3v}$ little point group, the states split as they transform according to the one-dimensional double representation  $\Gamma_{4}$ and $\Gamma_{5}$ from the $C_{3v}$ perspective. However, the  states transform according to the two-dimensional double-group representation $\Gamma_{6}$ and are thus do not split along this direction. Therefore, once band crossings arise between $\Gamma_{6}$ and $\Gamma_{4}$ and $\Gamma_{5}$, this will lead to a TDP, as shown in the insets in Figs. \ref{struc_bulk}(c) and \ref{struc_bulk}(d).

\begin{figure}
  \centering
  \includegraphics[width=3.5in]{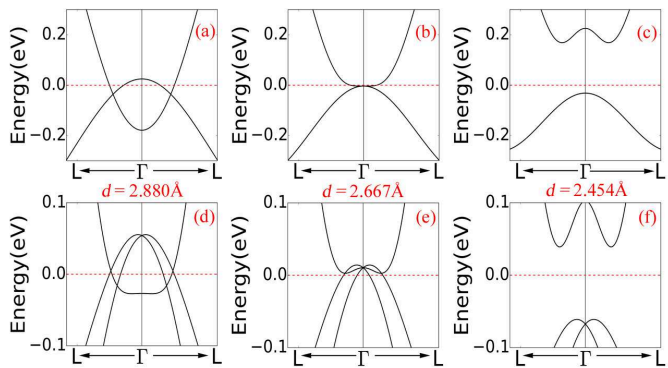}
  \caption{Band structure evolution of HfIrAs along L-$\Gamma$-L line with three different $d$ values. (a)(b)(c) are the cases in the absence of SOC and
  (d)(e)(f) are the cases in the presence of SOC, respectively. Note that the distances in (b) and (e) are the real crystal value.}\label{BIA_HfIrAs}
\end{figure}

\begin{figure}
  \centering
  \includegraphics[width=3.5in]{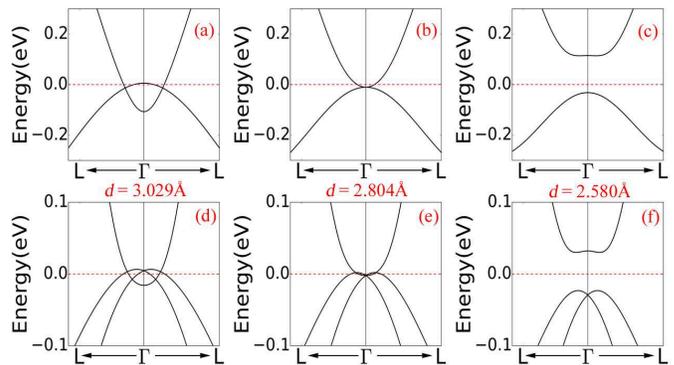}
  \caption{Band structure evolution of HfIrBi along L-$\Gamma$-L line as HfIrAs case.}\label{BIA_HfIrBi}
\end{figure}

It should be noted that the coupling between the $d$ states from the tetrahedrally nearest neighboring Hf and Ir atoms directly determine the formation of the $\Gamma_{8}$ band. Thus, the evolution of the band structure was investigated as a function of the relative distance between Hf and Ir along the [111] direction. Since varying the distance as described above does not violate the $C_{3v}$ symmetry, it can be considered to be a symmetry-allowed perturbation that drives the system from one phase to another and vice versa. Fig. \ref{BIA_HfIrAs} and Fig. \ref{BIA_HfIrBi} show the evolution of the band structure for HfIrAs and HfIrBi for three different distance values. When Hf and Ir atoms are close, the two bands are separated from each other due to the strong coupling between the $d$ states. As the distance increases, the coupling becomes weaker, and they begin to touch and anticross. In the presence of the SOC, each of the anticrossing points will split into a pair of TDPs. As the critical value of the distance is in the ideal structure, we expect that these two materials can appear on either side of the phase diagram under different synthesis conditions.

\begin{figure}
  \centering
  \includegraphics[width=3.5in,height=3.0in]{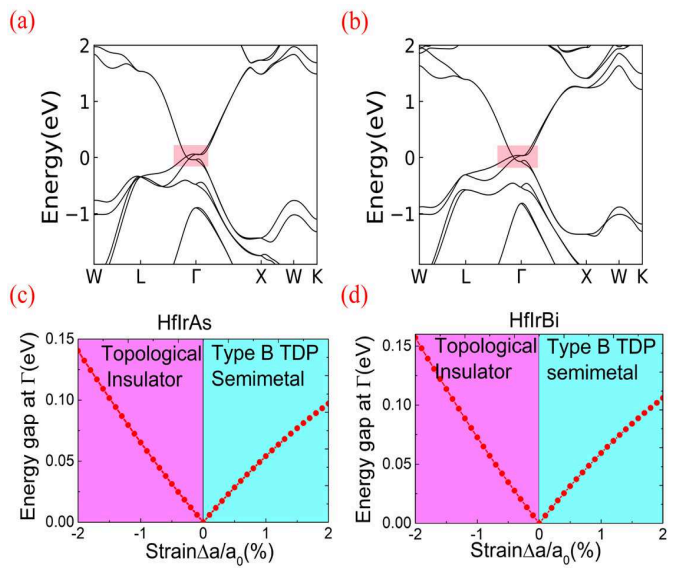}
  \caption{(a) and (b) are the band structure of HfIrAs and HfIrBi in the presence of $2\%$ tensile strain along $[111]$ direction. A band gap at the $\Gamma$ point is opened and a pairs of TDP along $\Gamma$-L exist. The additional TDP originates from the broken of $T_{d}$. (c) and (d) show the complete phase diagram of HfIrAs and HfIrBi as a function of strain. Under the tensile strain(compressive strain), the two TDP semimetals will enter the type B TDP sememetal phase(topological insulator phase).}\label{strain_ek}
\end{figure}

Next, we elucidate the origin of the TDP in terms of the effective Hamiltonian. A 4-band model \cite{PhysRev.100.580,voon2009kp} can be used to describe the band structure near the $\Gamma$ point according to the $k\cdot p$ theory. The effective Hamiltonian is:
\begin{equation}\label{HH}
       H=H_{0} + H_{BIA}
\end{equation}
     where $H_{0}$ denotes the Luttinger Hamiltonian and $H_{BIA}$ represents the BIA term due to the absence of inversion symmetry.
The details of the Luntinger Hamiltonian are as follows:
\begin{equation}\label{H0}
      H_{0}=\gamma _{1}k^{2}+\gamma
_{2}\sum _{i}k_{i}^{2}J_{i}^{2}+\gamma _{3}\left[
k_{x}k_{y} \left\{ J_{x},J_{y}\right\} +\mathrm{c.p.}\right]
    \end{equation}
    where $\gamma_{1},\gamma_{2},\gamma_{3}$ are the Luttinger parameters, $J_{i}$ denotes the spin-3/2 matrices,
    $\left\{ \right\}$ denotes anti-commutator and c.p. refers to cyclic permutations.
    The BIA Hamiltonian as the main perturbation is:
 \begin{equation}\label{HBIA}
       H_{BIA}=C\left[ k_{x}\left\{ J_{x},J_{y}^{2}-J_{z}^{2}\right\}
+\mathrm{c.p.}\right]
     \end{equation}

    Without BIA term, the band structure along the $[111]$ direction can be easily obtained by solving the Eq.(\ref{H0}): $E(k) = \left( \gamma _{1}+3\gamma _{2}\pm 3/4\sqrt{\gamma _{2}^{2}+\gamma_{3}^{2}}\right) k^{2} $ where $k_{x}=k_{y}=k_{z}=k$.
    The two bands are both two-degenerate and the TDP is not expected to be away from $\Gamma$. This is caused by the presence of inversion symmetry in Eq.(\ref{H0}) and is consistent with the Kramers degenerate theory.
    Nevertheless, with the BIA term, the spin splits along $[111]$ are $0,0,\pm \frac{3}{\sqrt{2}}Ck$. The parameter $C$ denotes the strength of BIA. The last band dispersion are $E_{\Gamma _{6}}(k) = \left( \gamma _{1}+3\gamma _{2}+3/4\sqrt{\gamma_{2}^{2}+\gamma _{3}^{2}}\right) k^{2}$ and $E_{\Gamma _{4,5}}(k) =\left( \gamma _{1}+3\gamma _{2}-3/4\sqrt{\gamma
_{2}^{2}+\gamma _{3}^{2}}\right) k^{2}\pm 3C/\sqrt{2}k$. As a result, the TDP position can be derived from $E_{\Gamma _{6}}(k)=E_{\Gamma _{4}}(k)$ or $E_{\Gamma _{6}}(k)=E_{\Gamma _{5}}(k)$: $\left( k_{T},k_{T},k_{T}\right)$ where $k_{T} =\pm \sqrt{\frac{2}{\gamma _{2}^{2}+\gamma _{3}^{2}}}C$. Therefore, although the magnitude of $C$ is small, it plays key role in realizing TDP in cubic space group. That is to say, the BIA is the source of the TDP in cubic crystal structures, such as III-V and half-Heusler materials. The example of the NaCu$_{3}$Te$_{2}$ family of TDP semimetals \cite{PhysRevLett.119.256402,PhysRevB.96.241204} also confirms this conclusion. The splitting of $\Gamma_{4}$ and $\Gamma_{5}$ is primarily due to the BIA effect \cite{voon2009kp} and it is universal for TDPs along $C_{3}$ axis.

Further, the strain can be considered to be an important band engineering tool that can be used to tune the band structure. According to the previous arguments \cite{PhysRevX.6.031003}, the case of an ideal HgTe is a trivial scenario while the case with tensile strain along the $[111]$ direction is a type B TDP. Hence, tensile strain was added along the $[111]$ direction for a fixed the cell volume  as this type of strain breaks the cubic symmetry and lifts the degenerate $\Gamma$ point. Consequently, a band gap appears at $\Gamma$ point. The band structure of the strained HfIrAs and HfIrBi are shown in Fig. \ref{strain_ek}(a) and \ref{strain_ek}(b).
Note that the number of TDP along $[111]$ direction is doubled. As there are $8$ $C_{3}$ axis, the total number of TDP now is $16$. This is due to the lifting at the $\Gamma$ point. This increases the number and changes the positions of TDPs.
Based on the above described effective Hamiltonian, the perturbation Hamiltonian induced by the strain is
   \begin{equation}\label{H_strain1}
     H_{strain}=\frac{\epsilon _{111}}{3}d\left[ \left\{ J_{x},J_{y}\right\} +\mathrm{c.p.}%
\right]
   \end{equation}
   where the $d$ is the strain elasticity parameter and the $\epsilon _{111}$ is determined by the strength of the strain. A value of $\epsilon _{111}>0$ represents tensile strain and $\epsilon _{111}<0$ represents compressive strain. Here we ignore the mixed terms between the strain $\epsilon$ and wavevector $k$ and retain the lowest order approximation. Under this approximation, the splitting energy at $\Gamma$ point is equal to $\left\vert d\epsilon _{111}\right\vert $. The above analysis
  indicates that the gap at the $\Gamma$ point increases linearly with the increasing strain. If the tensile strain is replaced by compressive strain, the system begins to enter the topological insulator phase as described in \cite{PhysRevB.87.045202}. The results in Fig. \ref{strain_ek}(c) and \ref{strain_ek}(d), which were calculated from first principle, support above theoretical analysis.

\begin{figure}
  \centering
  \includegraphics[width=3.5in]{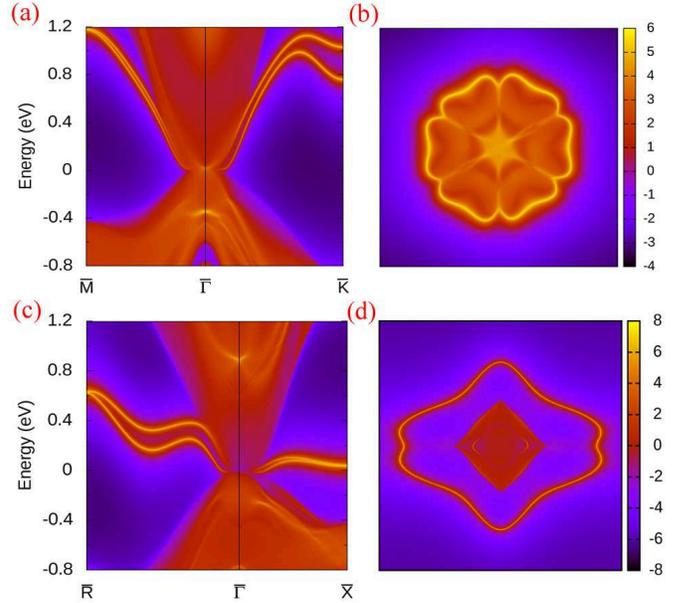}
  \caption{Local density of state(LDOS) of the surface and fermi arcs of HfIrAs. (a) and (c) are the LDOS for the type \uppercase\expandafter{\romannumeral1} and \uppercase\expandafter{\romannumeral2} surface. (b) and (d) are the fermi arcs for the type \uppercase\expandafter{\romannumeral1} and \uppercase\expandafter{\romannumeral2} surface. Here the energy level is set $E=0$eV.} \label{LDOS_HfIrAs}
\end{figure}
\begin{figure}
  \centering
  \includegraphics[width=3.5in]{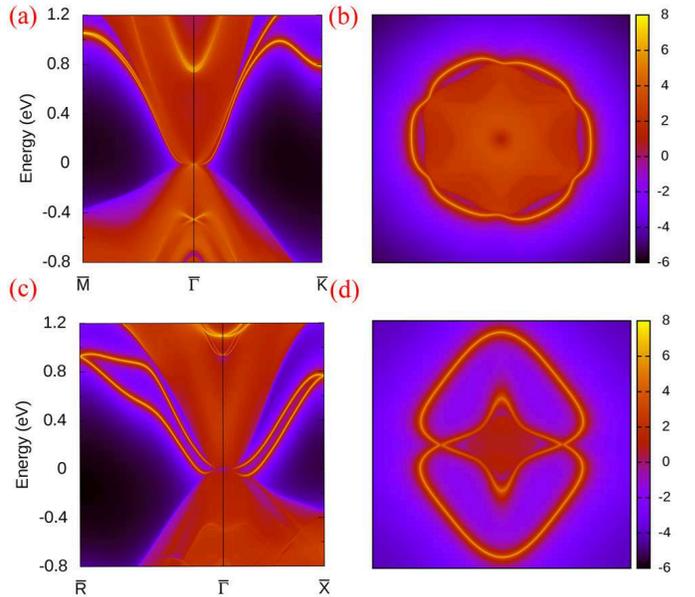}
  \caption{The LDOS of the surface and the fermi arcs of HfIrBi as Fig \ref{LDOS_HfIrAs}.} \label{LDOS_HfIrBi}
\end{figure}

\begin{figure}
  \centering
  \includegraphics[width=3.5in]{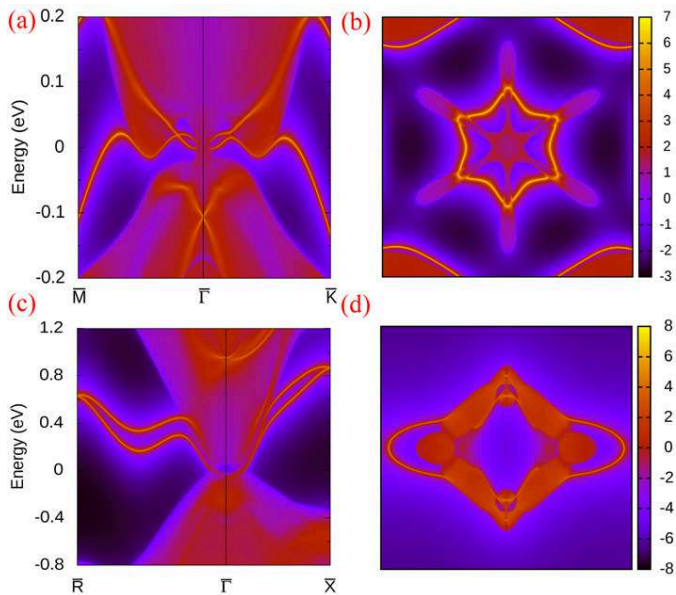}
  \caption{The LDOS of the surface and the fermi arcs of HfIrAs in the presence of $2\%$ tensile strain along
   $[111]$ direction.} \label{LDOS_strain}
\end{figure}

\begin{figure*}
  \centering
  \includegraphics[width=6.8in]{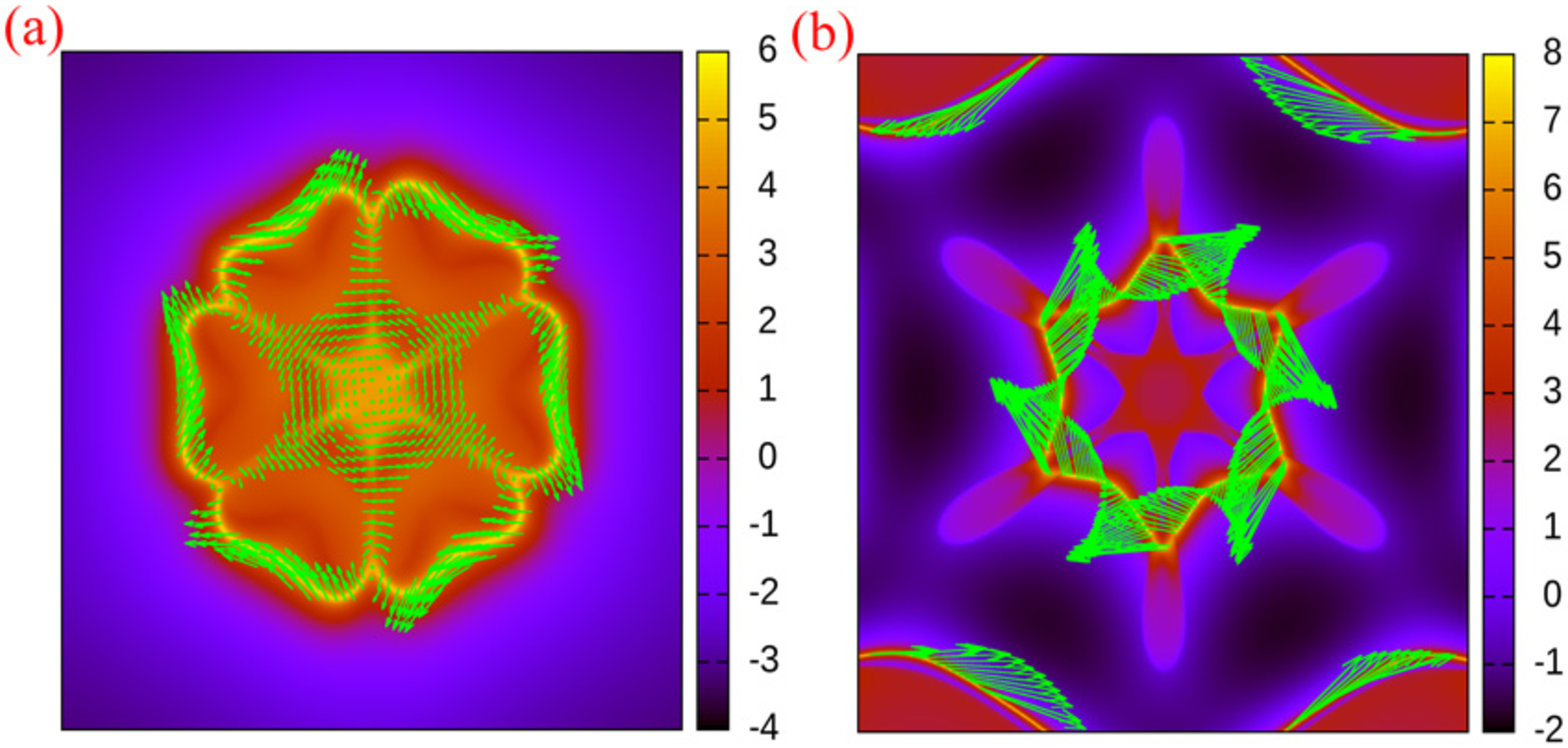}
  \caption{(a) and (b) are the spin texture of HfIrAs without and with tensile strain, respectively. }\label{spintexture}
\end{figure*}

Based on the unique properties of the bulk states, the corresponding surface states and fermi arcs may bring differences. As is known, surface states are used to characterize topological materials as they are easily detected in experiments. In topological insulator, gapless surface states are the focus. In contrast, in DSMs and WSMs, both the surface states and fermi arcs that link the Dirac and Weyl points are the focus. Since the TDP semimetals can be viewed as an intermediate phase between DSMs and WSMs \cite{PhysRevX.6.031003}, their surface states and fermi arcs are worthy of careful investigation. A key factor for surface states is the crystal face orientation. Here, we distinguish two types of crystal surface depending on the projected position of the $C_{3}$ axis.
For the type \uppercase\expandafter{\romannumeral1} surface $(111)$, six L points are projected to six different $\mathrm{\overline{M}}$ points, while the other two are projected to $\overline{\Gamma}$ point in the surface BZ. For type \uppercase\expandafter{\romannumeral2} surface $(001)$ or $(110)$, L points are projected to the corner of the square surface BZ. All of these projections from the bulk to surface BZ are illustrated in Fig. \ref{struc_bulk}(b).

  Since a typical TDP can be considered as two Weyl points with opposite charities, two fermi arcs are expected to emerge from a TDP.
  In the case of type \uppercase\expandafter{\romannumeral1}, two TDPs are projected at the $\overline{\Gamma}$ point and the other six TDPs are projected onto the line of $\overline{\Gamma}-\mathrm{\overline{M}}$. As a consequence, the surface bands disappear from the surface BZ boundary to the center $\overline{\Gamma}$ point. Meanwhile, the six TDPs are linked by the six fermi arcs as shown in Fig. \ref{LDOS_HfIrAs}(b). This pattern
  is preserved by the $C_{3}$ symmetry. These results agree well with those of the previous half-Heusler materials \cite{PhysRevLett.119.136401}.
  In the other case, no TDP is projected at $\overline{\Gamma}$ point on the surface BZ. Thus, the surface bands disappear at the TDPs
  As the $C_{2}$ point group preserves the surface BZ,
  it keeps the degenerate of the surface bands at $\mathrm{\overline{R}}$ and $\mathrm{\overline{X}}$ point and the resulting pattern of  the fermi arcs are different from the $(111)$ surface case.  This can be seen in Fig. \ref{LDOS_HfIrAs}(c) and \ref{LDOS_HfIrAs}(d). The corresponding surface states and fermi arcs of HfIrBi are shown in Fig. \ref{LDOS_HfIrBi}.

In the preceding discussion, the surface state and fermi arcs were obtained without tensile strain. Now, tensile strain is considered. Under the tensile strain along the $[111]$ direction, the tetrahedral point group $T_{d}$ is reduced to the trigonal point group $C_{3v}$, which has significant influence on the pattern of the fermi arcs. For Type \uppercase\expandafter{\romannumeral1}, the $C_{3}$ symmetry along the $[111]$ direction is maintained.
Thus, the fermi arcs should also retain a similar hexagon pattern and this is confirmed by Fig. \ref{LDOS_strain}(b). However, the linked lines between the center and the outer corners of the pattern, split into two branches due to the broken of $C_{3}$ symmetry along the other equivalent $[111]$ direction, which doubles the number of TDPs. Although the projected positions of the TDP pairs are very close, they are not identical. At the same time, no gap is observed at $\overline{\Gamma}$ points as the gapless projected TDPs at the $\overline{\Gamma}$ points hide the bulk gap.
For the type \uppercase\expandafter{\romannumeral1}, a band gap appears at the $\overline{\Gamma}$, as indicated by the blue area near the fermi level as there is no projected TDP at $\overline{\Gamma}$. These differences can be clear seen in Fig. \ref{LDOS_strain}.

Another special feature of topological material is the spin texture of the surface states. Little information is available on the spin texture of surface states, although more is known regarding the bulk states \cite{PhysRevB.87.045202} for TDP semimetal. In this study, the spin textures of the surface states were explored with and without tensile strain along the $[111]$ direction and the results are shown in Fig. \ref{spintexture}. The chiral spin texture was found to be widely distributed in the bright area delineated by the fermi arcs. This phenomenon is similar to that in topological insulator in which the spin texture most widely distributed on the gapless surface states. However, some differences do exist. The spin texture, disappears near the six corner of the fermi arcs as the bulk material \cite{PhysRevB.87.045202}, is flipped at these critical points due to the changes in the spin winding number, which is the result of bulk-edge correspondence.

\section{Methodology}
    The calculations were performed using the Vienna ab initio simulation package(VASP) \cite{kresse1996efficient}. The core-valence interaction was depicted by the projector augmented wave method \cite{blochl1994projector} and the generalized gradient approximation of Predew-Burke-Ernzerhof(GGA-PBE) \cite{PhysRevLett.77.3865} was selected for the
    exchange-correlation functional. The kinetic energy cutoff was set to 560eV
    and the k-point mesh was $9\times9\times9$. SOC was included in the band calculation. The lattice constant of HfIrAs and HfIrBi were $a=6.159${\AA} and $a=6.476${\AA} \cite{gautier2015prediction} and the $4c$ $4b$ and $4a$ Wyckoff positions were occupied by the Hf, Ir and As(Bi) atom. In order to simulate the strain, the lattice constant was varied  for a fixed the cell volume.
    The underestimation of the band gap was improved via the Heyd-Scuseria-Ernzerhof hybrid functional described in the band structure calculation in \cite{gautier2015prediction}. In addition, the unique surface states and fermi arcs, were visualized using the WannierTools package \cite{WU2017}. The numerical
    tight-binding model was constructed on the basis of maximally localized Wannier functions \cite{mostofi2008wannier90}.

\section{Conclusion}
In conclusion, the existence of two ideal TDP semimetals, HfIrAs and HfIrBi, by first principle calculations. The origin of the TDP is revealed via the frame of energy band evolution at atomic level and 4-band ${k \cdot p}$ effective Hamiltonian. The BIA performed an essential role in the formation of TDP in cubic crystals. In addition, the influence of the strain on the band structure was investigated. The presence of tensile strain along the $[111]$ direction induces a gap at the $\Gamma$ point and doubles the number of TDP. As was described, surface states, fermi arcs and spin texture for two different type of crystal face orientation were investigated. The hexagon pattern of the fermi arcs was also observed to be consistent with those of previous predicted half-Heusler TDP semimetal \cite{PhysRevLett.119.136401}. At the same time, the spin texture, that was found to be distributed mostly on the fermi arcs, disappeared at some critical points in the surface states.
These observable phenomena can be detected quite easily in experiments as there is no other quasiparticle band around the TDPs. Thus, the two materials provide a very useful platform on which to study the electron properties of TDPs semimetal, especially in terms of their quantum transport properties.

\section{Acknowledgments}
This work was financially supported
by the ERC Advanced Grant No. 291472 ¡®Idea Heusler,¡¯
ERC Advanced Grant No. 742068¨CTOPMAT and Deutsche
ForschungsgemeinschaftDFGunderSFB1143.
\bibliographystyle{apsrev}
\bibliography{TDP}

\begin{thebibliography}{34}
\expandafter\ifx\csname natexlab\endcsname\relax\def\natexlab#1{#1}\fi
\expandafter\ifx\csname bibnamefont\endcsname\relax
  \def\bibnamefont#1{#1}\fi
\expandafter\ifx\csname bibfnamefont\endcsname\relax
  \def\bibfnamefont#1{#1}\fi
\expandafter\ifx\csname citenamefont\endcsname\relax
  \def\citenamefont#1{#1}\fi
\expandafter\ifx\csname url\endcsname\relax
  \def\url#1{\texttt{#1}}\fi
\expandafter\ifx\csname urlprefix\endcsname\relax\def\urlprefix{URL }\fi
\providecommand{\bibinfo}[2]{#2}
\providecommand{\eprint}[2][]{\url{#2}}

\bibitem[{\citenamefont{K{\"o}nig et~al.}(2007)\citenamefont{K{\"o}nig,
  Wiedmann, Br{\"u}ne, Roth, Buhmann, Molenkamp, Qi, and
  Zhang}}]{konig2007quantum}
\bibinfo{author}{\bibfnamefont{M.}~\bibnamefont{K{\"o}nig}},
  \bibinfo{author}{\bibfnamefont{S.}~\bibnamefont{Wiedmann}},
  \bibinfo{author}{\bibfnamefont{C.}~\bibnamefont{Br{\"u}ne}},
  \bibinfo{author}{\bibfnamefont{A.}~\bibnamefont{Roth}},
  \bibinfo{author}{\bibfnamefont{H.}~\bibnamefont{Buhmann}},
  \bibinfo{author}{\bibfnamefont{L.~W.} \bibnamefont{Molenkamp}},
  \bibinfo{author}{\bibfnamefont{X.-L.} \bibnamefont{Qi}}, \bibnamefont{and}
  \bibinfo{author}{\bibfnamefont{S.-C.} \bibnamefont{Zhang}},
  \bibinfo{journal}{Science} \textbf{\bibinfo{volume}{318}},
  \bibinfo{pages}{766} (\bibinfo{year}{2007}).

\bibitem[{\citenamefont{Liu et~al.}(2008)\citenamefont{Liu, Hughes, Qi, Wang,
  and Zhang}}]{liu2008quantum}
\bibinfo{author}{\bibfnamefont{C.}~\bibnamefont{Liu}},
  \bibinfo{author}{\bibfnamefont{T.~L.} \bibnamefont{Hughes}},
  \bibinfo{author}{\bibfnamefont{X.-L.} \bibnamefont{Qi}},
  \bibinfo{author}{\bibfnamefont{K.}~\bibnamefont{Wang}}, \bibnamefont{and}
  \bibinfo{author}{\bibfnamefont{S.-C.} \bibnamefont{Zhang}},
  \bibinfo{journal}{Phys. Rev. Lett.} \textbf{\bibinfo{volume}{100}},
  \bibinfo{pages}{236601} (\bibinfo{year}{2008}).

\bibitem[{\citenamefont{Hasan and Kane}(2010)}]{hasan2010colloquium}
\bibinfo{author}{\bibfnamefont{M.~Z.} \bibnamefont{Hasan}} \bibnamefont{and}
  \bibinfo{author}{\bibfnamefont{C.~L.} \bibnamefont{Kane}},
  \bibinfo{journal}{Rev. Mod. Phys.} \textbf{\bibinfo{volume}{82}},
  \bibinfo{pages}{3045} (\bibinfo{year}{2010}).

\bibitem[{\citenamefont{Qi and Zhang}(2011)}]{qi2011topological}
\bibinfo{author}{\bibfnamefont{X.-L.} \bibnamefont{Qi}} \bibnamefont{and}
  \bibinfo{author}{\bibfnamefont{S.-C.} \bibnamefont{Zhang}},
  \bibinfo{journal}{Rev. Mod. Phys.} \textbf{\bibinfo{volume}{83}},
  \bibinfo{pages}{1057} (\bibinfo{year}{2011}).

\bibitem[{\citenamefont{Fu}(2011)}]{fu2011topological}
\bibinfo{author}{\bibfnamefont{L.}~\bibnamefont{Fu}}, \bibinfo{journal}{Phys.
  Rev. Lett.} \textbf{\bibinfo{volume}{106}}, \bibinfo{pages}{106802}
  (\bibinfo{year}{2011}).

\bibitem[{\citenamefont{Tanaka et~al.}(2012)\citenamefont{Tanaka, Ren, Sato,
  Nakayama, Souma, Takahashi, Segawa, and Ando}}]{tanaka2012experimental}
\bibinfo{author}{\bibfnamefont{Y.}~\bibnamefont{Tanaka}},
  \bibinfo{author}{\bibfnamefont{Z.}~\bibnamefont{Ren}},
  \bibinfo{author}{\bibfnamefont{T.}~\bibnamefont{Sato}},
  \bibinfo{author}{\bibfnamefont{K.}~\bibnamefont{Nakayama}},
  \bibinfo{author}{\bibfnamefont{S.}~\bibnamefont{Souma}},
  \bibinfo{author}{\bibfnamefont{T.}~\bibnamefont{Takahashi}},
  \bibinfo{author}{\bibfnamefont{K.}~\bibnamefont{Segawa}}, \bibnamefont{and}
  \bibinfo{author}{\bibfnamefont{Y.}~\bibnamefont{Ando}},
  \bibinfo{journal}{Nat. Phys.} \textbf{\bibinfo{volume}{8}},
  \bibinfo{pages}{800} (\bibinfo{year}{2012}).

\bibitem[{\citenamefont{Liu et~al.}(2014)\citenamefont{Liu, Zhou, Zhang, Wang,
  Weng, Prabhakaran, Mo, Shen, Fang, Dai et~al.}}]{Liu864}
\bibinfo{author}{\bibfnamefont{Z.~K.} \bibnamefont{Liu}},
  \bibinfo{author}{\bibfnamefont{B.}~\bibnamefont{Zhou}},
  \bibinfo{author}{\bibfnamefont{Y.}~\bibnamefont{Zhang}},
  \bibinfo{author}{\bibfnamefont{Z.~J.} \bibnamefont{Wang}},
  \bibinfo{author}{\bibfnamefont{H.~M.} \bibnamefont{Weng}},
  \bibinfo{author}{\bibfnamefont{D.}~\bibnamefont{Prabhakaran}},
  \bibinfo{author}{\bibfnamefont{S.-K.} \bibnamefont{Mo}},
  \bibinfo{author}{\bibfnamefont{Z.~X.} \bibnamefont{Shen}},
  \bibinfo{author}{\bibfnamefont{Z.}~\bibnamefont{Fang}},
  \bibinfo{author}{\bibfnamefont{X.}~\bibnamefont{Dai}}, \bibnamefont{et~al.},
  \bibinfo{journal}{Science} \textbf{\bibinfo{volume}{343}},
  \bibinfo{pages}{864} (\bibinfo{year}{2014}).

\bibitem[{\citenamefont{Wang et~al.}(2012)\citenamefont{Wang, Sun, Chen,
  Franchini, Xu, Weng, Dai, and Fang}}]{PhysRevB.85.195320}
\bibinfo{author}{\bibfnamefont{Z.}~\bibnamefont{Wang}},
  \bibinfo{author}{\bibfnamefont{Y.}~\bibnamefont{Sun}},
  \bibinfo{author}{\bibfnamefont{X.-Q.} \bibnamefont{Chen}},
  \bibinfo{author}{\bibfnamefont{C.}~\bibnamefont{Franchini}},
  \bibinfo{author}{\bibfnamefont{G.}~\bibnamefont{Xu}},
  \bibinfo{author}{\bibfnamefont{H.}~\bibnamefont{Weng}},
  \bibinfo{author}{\bibfnamefont{X.}~\bibnamefont{Dai}}, \bibnamefont{and}
  \bibinfo{author}{\bibfnamefont{Z.}~\bibnamefont{Fang}},
  \bibinfo{journal}{Phys. Rev. B} \textbf{\bibinfo{volume}{85}},
  \bibinfo{pages}{195320} (\bibinfo{year}{2012}).

\bibitem[{\citenamefont{Neupane et~al.}(2014)\citenamefont{Neupane, Xu, Sankar,
  Alidoust, Bian, Liu, Belopolski, Chang, Jeng, Lin
  et~al.}}]{neupane2014observation}
\bibinfo{author}{\bibfnamefont{M.}~\bibnamefont{Neupane}},
  \bibinfo{author}{\bibfnamefont{S.-Y.} \bibnamefont{Xu}},
  \bibinfo{author}{\bibfnamefont{R.}~\bibnamefont{Sankar}},
  \bibinfo{author}{\bibfnamefont{N.}~\bibnamefont{Alidoust}},
  \bibinfo{author}{\bibfnamefont{G.}~\bibnamefont{Bian}},
  \bibinfo{author}{\bibfnamefont{C.}~\bibnamefont{Liu}},
  \bibinfo{author}{\bibfnamefont{I.}~\bibnamefont{Belopolski}},
  \bibinfo{author}{\bibfnamefont{T.-R.} \bibnamefont{Chang}},
  \bibinfo{author}{\bibfnamefont{H.-T.} \bibnamefont{Jeng}},
  \bibinfo{author}{\bibfnamefont{H.}~\bibnamefont{Lin}}, \bibnamefont{et~al.},
  \bibinfo{journal}{Nat. Commun.} \textbf{\bibinfo{volume}{5}},
  \bibinfo{pages}{3786} (\bibinfo{year}{2014}).

\bibitem[{\citenamefont{Burkov and Balents}(2011)}]{burkov2011weyl}
\bibinfo{author}{\bibfnamefont{A.}~\bibnamefont{Burkov}} \bibnamefont{and}
  \bibinfo{author}{\bibfnamefont{L.}~\bibnamefont{Balents}},
  \bibinfo{journal}{Phys. Rev. Lett.} \textbf{\bibinfo{volume}{107}},
  \bibinfo{pages}{127205} (\bibinfo{year}{2011}).

\bibitem[{\citenamefont{Wan et~al.}(2011)\citenamefont{Wan, Turner, Vishwanath,
  and Savrasov}}]{wan2011topological}
\bibinfo{author}{\bibfnamefont{X.}~\bibnamefont{Wan}},
  \bibinfo{author}{\bibfnamefont{A.~M.} \bibnamefont{Turner}},
  \bibinfo{author}{\bibfnamefont{A.}~\bibnamefont{Vishwanath}},
  \bibnamefont{and} \bibinfo{author}{\bibfnamefont{S.~Y.}
  \bibnamefont{Savrasov}}, \bibinfo{journal}{Phys. Rev. B}
  \textbf{\bibinfo{volume}{83}}, \bibinfo{pages}{205101}
  (\bibinfo{year}{2011}).

\bibitem[{\citenamefont{Weng et~al.}(2015)\citenamefont{Weng, Fang, Fang,
  Bernevig, and Dai}}]{weng2015weyl}
\bibinfo{author}{\bibfnamefont{H.}~\bibnamefont{Weng}},
  \bibinfo{author}{\bibfnamefont{C.}~\bibnamefont{Fang}},
  \bibinfo{author}{\bibfnamefont{Z.}~\bibnamefont{Fang}},
  \bibinfo{author}{\bibfnamefont{B.~A.} \bibnamefont{Bernevig}},
  \bibnamefont{and} \bibinfo{author}{\bibfnamefont{X.}~\bibnamefont{Dai}},
  \bibinfo{journal}{Phys. Rev. X} \textbf{\bibinfo{volume}{5}},
  \bibinfo{pages}{011029} (\bibinfo{year}{2015}).

\bibitem[{\citenamefont{Lv et~al.}(2015)\citenamefont{Lv, Weng, Fu, Wang, Miao,
  Ma, Richard, Huang, Zhao, Chen et~al.}}]{lv2015experimental}
\bibinfo{author}{\bibfnamefont{B.~Q.} \bibnamefont{Lv}},
  \bibinfo{author}{\bibfnamefont{H.~M.} \bibnamefont{Weng}},
  \bibinfo{author}{\bibfnamefont{B.~B.} \bibnamefont{Fu}},
  \bibinfo{author}{\bibfnamefont{X.~P.} \bibnamefont{Wang}},
  \bibinfo{author}{\bibfnamefont{H.}~\bibnamefont{Miao}},
  \bibinfo{author}{\bibfnamefont{J.}~\bibnamefont{Ma}},
  \bibinfo{author}{\bibfnamefont{P.}~\bibnamefont{Richard}},
  \bibinfo{author}{\bibfnamefont{X.~C.} \bibnamefont{Huang}},
  \bibinfo{author}{\bibfnamefont{L.~X.} \bibnamefont{Zhao}},
  \bibinfo{author}{\bibfnamefont{G.~F.} \bibnamefont{Chen}},
  \bibnamefont{et~al.}, \bibinfo{journal}{Phys. Rev. X}
  \textbf{\bibinfo{volume}{5}}, \bibinfo{pages}{031013} (\bibinfo{year}{2015}).

\bibitem[{\citenamefont{Bradlyn et~al.}(2016)\citenamefont{Bradlyn, Cano, Wang,
  Vergniory, Felser, Cava, and Bernevig}}]{bradlyn2016beyond}
\bibinfo{author}{\bibfnamefont{B.}~\bibnamefont{Bradlyn}},
  \bibinfo{author}{\bibfnamefont{J.}~\bibnamefont{Cano}},
  \bibinfo{author}{\bibfnamefont{Z.}~\bibnamefont{Wang}},
  \bibinfo{author}{\bibfnamefont{M.}~\bibnamefont{Vergniory}},
  \bibinfo{author}{\bibfnamefont{C.}~\bibnamefont{Felser}},
  \bibinfo{author}{\bibfnamefont{R.}~\bibnamefont{Cava}}, \bibnamefont{and}
  \bibinfo{author}{\bibfnamefont{B.~A.} \bibnamefont{Bernevig}},
  \bibinfo{journal}{Science} \textbf{\bibinfo{volume}{353}},
  \bibinfo{pages}{aaf5037} (\bibinfo{year}{2016}).

\bibitem[{\citenamefont{Zaheer et~al.}(2013)\citenamefont{Zaheer, Young,
  Cellucci, Teo, Kane, Mele, and Rappe}}]{PhysRevB.87.045202}
\bibinfo{author}{\bibfnamefont{S.}~\bibnamefont{Zaheer}},
  \bibinfo{author}{\bibfnamefont{S.~M.} \bibnamefont{Young}},
  \bibinfo{author}{\bibfnamefont{D.}~\bibnamefont{Cellucci}},
  \bibinfo{author}{\bibfnamefont{J.~C.~Y.} \bibnamefont{Teo}},
  \bibinfo{author}{\bibfnamefont{C.~L.} \bibnamefont{Kane}},
  \bibinfo{author}{\bibfnamefont{E.~J.} \bibnamefont{Mele}}, \bibnamefont{and}
  \bibinfo{author}{\bibfnamefont{A.~M.} \bibnamefont{Rappe}},
  \bibinfo{journal}{Phys. Rev. B} \textbf{\bibinfo{volume}{87}},
  \bibinfo{pages}{045202} (\bibinfo{year}{2013}).

\bibitem[{\citenamefont{Winkler et~al.}(2016)\citenamefont{Winkler, Wu, Troyer,
  Krogstrup, and Soluyanov}}]{PhysRevLett.117.076403}
\bibinfo{author}{\bibfnamefont{G.~W.} \bibnamefont{Winkler}},
  \bibinfo{author}{\bibfnamefont{Q.}~\bibnamefont{Wu}},
  \bibinfo{author}{\bibfnamefont{M.}~\bibnamefont{Troyer}},
  \bibinfo{author}{\bibfnamefont{P.}~\bibnamefont{Krogstrup}},
  \bibnamefont{and} \bibinfo{author}{\bibfnamefont{A.~A.}
  \bibnamefont{Soluyanov}}, \bibinfo{journal}{Phys. Rev. Lett.}
  \textbf{\bibinfo{volume}{117}}, \bibinfo{pages}{076403}
  (\bibinfo{year}{2016}).

\bibitem[{\citenamefont{Zhu et~al.}(2016)\citenamefont{Zhu, Winkler, Wu, Li,
  and Soluyanov}}]{PhysRevX.6.031003}
\bibinfo{author}{\bibfnamefont{Z.}~\bibnamefont{Zhu}},
  \bibinfo{author}{\bibfnamefont{G.~W.} \bibnamefont{Winkler}},
  \bibinfo{author}{\bibfnamefont{Q.}~\bibnamefont{Wu}},
  \bibinfo{author}{\bibfnamefont{J.}~\bibnamefont{Li}}, \bibnamefont{and}
  \bibinfo{author}{\bibfnamefont{A.~A.} \bibnamefont{Soluyanov}},
  \bibinfo{journal}{Phys. Rev. X} \textbf{\bibinfo{volume}{6}},
  \bibinfo{pages}{031003} (\bibinfo{year}{2016}).

\bibitem[{\citenamefont{Lv et~al.}(2017)\citenamefont{Lv, Feng, Xu, Gao, Ma,
  Kong, Richard, Huang, Strocov, Fang et~al.}}]{lv2017observation}
\bibinfo{author}{\bibfnamefont{B.}~\bibnamefont{Lv}},
  \bibinfo{author}{\bibfnamefont{Z.-L.} \bibnamefont{Feng}},
  \bibinfo{author}{\bibfnamefont{Q.-N.} \bibnamefont{Xu}},
  \bibinfo{author}{\bibfnamefont{X.}~\bibnamefont{Gao}},
  \bibinfo{author}{\bibfnamefont{J.-Z.} \bibnamefont{Ma}},
  \bibinfo{author}{\bibfnamefont{L.-Y.} \bibnamefont{Kong}},
  \bibinfo{author}{\bibfnamefont{P.}~\bibnamefont{Richard}},
  \bibinfo{author}{\bibfnamefont{Y.-B.} \bibnamefont{Huang}},
  \bibinfo{author}{\bibfnamefont{V.}~\bibnamefont{Strocov}},
  \bibinfo{author}{\bibfnamefont{C.}~\bibnamefont{Fang}}, \bibnamefont{et~al.},
  \bibinfo{journal}{Nature} \textbf{\bibinfo{volume}{546}},
  \bibinfo{pages}{627} (\bibinfo{year}{2017}).

\bibitem[{\citenamefont{Weng et~al.}(2016{\natexlab{a}})\citenamefont{Weng,
  Fang, Fang, and Dai}}]{PhysRevB.93.241202}
\bibinfo{author}{\bibfnamefont{H.}~\bibnamefont{Weng}},
  \bibinfo{author}{\bibfnamefont{C.}~\bibnamefont{Fang}},
  \bibinfo{author}{\bibfnamefont{Z.}~\bibnamefont{Fang}}, \bibnamefont{and}
  \bibinfo{author}{\bibfnamefont{X.}~\bibnamefont{Dai}},
  \bibinfo{journal}{Phys. Rev. B} \textbf{\bibinfo{volume}{93}},
  \bibinfo{pages}{241202} (\bibinfo{year}{2016}{\natexlab{a}}).

\bibitem[{\citenamefont{Weng et~al.}(2016{\natexlab{b}})\citenamefont{Weng,
  Fang, Fang, and Dai}}]{PhysRevB.94.165201}
\bibinfo{author}{\bibfnamefont{H.}~\bibnamefont{Weng}},
  \bibinfo{author}{\bibfnamefont{C.}~\bibnamefont{Fang}},
  \bibinfo{author}{\bibfnamefont{Z.}~\bibnamefont{Fang}}, \bibnamefont{and}
  \bibinfo{author}{\bibfnamefont{X.}~\bibnamefont{Dai}},
  \bibinfo{journal}{Phys. Rev. B} \textbf{\bibinfo{volume}{94}},
  \bibinfo{pages}{165201} (\bibinfo{year}{2016}{\natexlab{b}}).

\bibitem[{\citenamefont{Yu et~al.}(2017)\citenamefont{Yu, Yan, and
  Liu}}]{PhysRevB.95.235158}
\bibinfo{author}{\bibfnamefont{J.}~\bibnamefont{Yu}},
  \bibinfo{author}{\bibfnamefont{B.}~\bibnamefont{Yan}}, \bibnamefont{and}
  \bibinfo{author}{\bibfnamefont{C.-X.} \bibnamefont{Liu}},
  \bibinfo{journal}{Phys. Rev. B} \textbf{\bibinfo{volume}{95}},
  \bibinfo{pages}{235158} (\bibinfo{year}{2017}).

\bibitem[{\citenamefont{Fulga and Stern}(2017)}]{PhysRevB.95.241116}
\bibinfo{author}{\bibfnamefont{I.~C.} \bibnamefont{Fulga}} \bibnamefont{and}
  \bibinfo{author}{\bibfnamefont{A.}~\bibnamefont{Stern}},
  \bibinfo{journal}{Phys. Rev. B} \textbf{\bibinfo{volume}{95}},
  \bibinfo{pages}{241116} (\bibinfo{year}{2017}).

\bibitem[{\citenamefont{Yang et~al.}(2017)\citenamefont{Yang, Yu, Parkin,
  Felser, Liu, and Yan}}]{PhysRevLett.119.136401}
\bibinfo{author}{\bibfnamefont{H.}~\bibnamefont{Yang}},
  \bibinfo{author}{\bibfnamefont{J.}~\bibnamefont{Yu}},
  \bibinfo{author}{\bibfnamefont{S.~S.~P.} \bibnamefont{Parkin}},
  \bibinfo{author}{\bibfnamefont{C.}~\bibnamefont{Felser}},
  \bibinfo{author}{\bibfnamefont{C.-X.} \bibnamefont{Liu}}, \bibnamefont{and}
  \bibinfo{author}{\bibfnamefont{B.}~\bibnamefont{Yan}},
  \bibinfo{journal}{Phys. Rev. Lett.} \textbf{\bibinfo{volume}{119}},
  \bibinfo{pages}{136401} (\bibinfo{year}{2017}).

\bibitem[{\citenamefont{Xia and Li}(2017)}]{PhysRevB.96.241204}
\bibinfo{author}{\bibfnamefont{Y.}~\bibnamefont{Xia}} \bibnamefont{and}
  \bibinfo{author}{\bibfnamefont{G.}~\bibnamefont{Li}}, \bibinfo{journal}{Phys.
  Rev. B} \textbf{\bibinfo{volume}{96}}, \bibinfo{pages}{241204}
  (\bibinfo{year}{2017}).

\bibitem[{\citenamefont{Wang et~al.}(2017)\citenamefont{Wang, Sui, Shi, Pan,
  Zhang, Liu, Wei, Yan, and Huang}}]{PhysRevLett.119.256402}
\bibinfo{author}{\bibfnamefont{J.}~\bibnamefont{Wang}},
  \bibinfo{author}{\bibfnamefont{X.}~\bibnamefont{Sui}},
  \bibinfo{author}{\bibfnamefont{W.}~\bibnamefont{Shi}},
  \bibinfo{author}{\bibfnamefont{J.}~\bibnamefont{Pan}},
  \bibinfo{author}{\bibfnamefont{S.}~\bibnamefont{Zhang}},
  \bibinfo{author}{\bibfnamefont{F.}~\bibnamefont{Liu}},
  \bibinfo{author}{\bibfnamefont{S.-H.} \bibnamefont{Wei}},
  \bibinfo{author}{\bibfnamefont{Q.}~\bibnamefont{Yan}}, \bibnamefont{and}
  \bibinfo{author}{\bibfnamefont{B.}~\bibnamefont{Huang}},
  \bibinfo{journal}{Phys. Rev. Lett.} \textbf{\bibinfo{volume}{119}},
  \bibinfo{pages}{256402} (\bibinfo{year}{2017}).

\bibitem[{\citenamefont{Dresselhaus}(1955)}]{PhysRev.100.580}
\bibinfo{author}{\bibfnamefont{G.}~\bibnamefont{Dresselhaus}},
  \bibinfo{journal}{Phys. Rev.} \textbf{\bibinfo{volume}{100}},
  \bibinfo{pages}{580} (\bibinfo{year}{1955}).

\bibitem[{\citenamefont{Chadov et~al.}(2010)\citenamefont{Chadov, Qi,
  K{\"u}bler, Fecher, Felser, and Zhang}}]{chadov2010tunable}
\bibinfo{author}{\bibfnamefont{S.}~\bibnamefont{Chadov}},
  \bibinfo{author}{\bibfnamefont{X.}~\bibnamefont{Qi}},
  \bibinfo{author}{\bibfnamefont{J.}~\bibnamefont{K{\"u}bler}},
  \bibinfo{author}{\bibfnamefont{G.~H.} \bibnamefont{Fecher}},
  \bibinfo{author}{\bibfnamefont{C.}~\bibnamefont{Felser}}, \bibnamefont{and}
  \bibinfo{author}{\bibfnamefont{S.~C.} \bibnamefont{Zhang}},
  \bibinfo{journal}{Nat. Mater.} \textbf{\bibinfo{volume}{9}},
  \bibinfo{pages}{541} (\bibinfo{year}{2010}).

\bibitem[{\citenamefont{Gautier et~al.}(2015)\citenamefont{Gautier, Zhang, Hu,
  Yu, Lin, Sunde, Chon, Poeppelmeier, and Zunger}}]{gautier2015prediction}
\bibinfo{author}{\bibfnamefont{R.}~\bibnamefont{Gautier}},
  \bibinfo{author}{\bibfnamefont{X.}~\bibnamefont{Zhang}},
  \bibinfo{author}{\bibfnamefont{L.}~\bibnamefont{Hu}},
  \bibinfo{author}{\bibfnamefont{L.}~\bibnamefont{Yu}},
  \bibinfo{author}{\bibfnamefont{Y.}~\bibnamefont{Lin}},
  \bibinfo{author}{\bibfnamefont{T.~O.} \bibnamefont{Sunde}},
  \bibinfo{author}{\bibfnamefont{D.}~\bibnamefont{Chon}},
  \bibinfo{author}{\bibfnamefont{K.~R.} \bibnamefont{Poeppelmeier}},
  \bibnamefont{and} \bibinfo{author}{\bibfnamefont{A.}~\bibnamefont{Zunger}},
  \bibinfo{journal}{Nat. Chem.} \textbf{\bibinfo{volume}{7}},
  \bibinfo{pages}{308} (\bibinfo{year}{2015}).

\bibitem[{\citenamefont{Voon and Willatzen}(2009)}]{voon2009kp}
\bibinfo{author}{\bibfnamefont{L.~C. L.~Y.} \bibnamefont{Voon}}
  \bibnamefont{and}
  \bibinfo{author}{\bibfnamefont{M.}~\bibnamefont{Willatzen}},
  \emph{\bibinfo{title}{The kp method: electronic properties of
  semiconductors}} (\bibinfo{publisher}{Springer Science \& Business Media},
  \bibinfo{year}{2009}).

\bibitem[{\citenamefont{Kresse and
  Furthm{\"u}ller}(1996)}]{kresse1996efficient}
\bibinfo{author}{\bibfnamefont{G.}~\bibnamefont{Kresse}} \bibnamefont{and}
  \bibinfo{author}{\bibfnamefont{J.}~\bibnamefont{Furthm{\"u}ller}},
  \bibinfo{journal}{Phys. Rev. B} \textbf{\bibinfo{volume}{54}},
  \bibinfo{pages}{11169} (\bibinfo{year}{1996}).

\bibitem[{\citenamefont{Bl{\"o}chl}(1994)}]{blochl1994projector}
\bibinfo{author}{\bibfnamefont{P.~E.} \bibnamefont{Bl{\"o}chl}},
  \bibinfo{journal}{Phys. Rev. B} \textbf{\bibinfo{volume}{50}},
  \bibinfo{pages}{17953} (\bibinfo{year}{1994}).

\bibitem[{\citenamefont{Perdew et~al.}(1996)\citenamefont{Perdew, Burke, and
  Ernzerhof}}]{PhysRevLett.77.3865}
\bibinfo{author}{\bibfnamefont{J.~P.} \bibnamefont{Perdew}},
  \bibinfo{author}{\bibfnamefont{K.}~\bibnamefont{Burke}}, \bibnamefont{and}
  \bibinfo{author}{\bibfnamefont{M.}~\bibnamefont{Ernzerhof}},
  \bibinfo{journal}{Phys. Rev. Lett.} \textbf{\bibinfo{volume}{77}},
  \bibinfo{pages}{3865} (\bibinfo{year}{1996}).

\bibitem[{\citenamefont{Wu et~al.}(2018)\citenamefont{Wu, Zhang, Song, Troyer,
  and Soluyanov}}]{WU2017}
\bibinfo{author}{\bibfnamefont{Q.}~\bibnamefont{Wu}},
  \bibinfo{author}{\bibfnamefont{S.}~\bibnamefont{Zhang}},
  \bibinfo{author}{\bibfnamefont{H.-F.} \bibnamefont{Song}},
  \bibinfo{author}{\bibfnamefont{M.}~\bibnamefont{Troyer}}, \bibnamefont{and}
  \bibinfo{author}{\bibfnamefont{A.~A.} \bibnamefont{Soluyanov}},
  \bibinfo{journal}{Comput. Phys. Commun.} \textbf{\bibinfo{volume}{224}},
  \bibinfo{pages}{405 } (\bibinfo{year}{2018}), ISSN \bibinfo{issn}{0010-4655}.

\bibitem[{\citenamefont{Mostofi et~al.}(2008)\citenamefont{Mostofi, Yates, Lee,
  Souza, Vanderbilt, and Marzari}}]{mostofi2008wannier90}
\bibinfo{author}{\bibfnamefont{A.~A.} \bibnamefont{Mostofi}},
  \bibinfo{author}{\bibfnamefont{J.~R.} \bibnamefont{Yates}},
  \bibinfo{author}{\bibfnamefont{Y.-S.} \bibnamefont{Lee}},
  \bibinfo{author}{\bibfnamefont{I.}~\bibnamefont{Souza}},
  \bibinfo{author}{\bibfnamefont{D.}~\bibnamefont{Vanderbilt}},
  \bibnamefont{and} \bibinfo{author}{\bibfnamefont{N.}~\bibnamefont{Marzari}},
  \bibinfo{journal}{Comput. Phys. Commun.} \textbf{\bibinfo{volume}{178}},
  \bibinfo{pages}{685} (\bibinfo{year}{2008}).

\end{thebibliography}


\end{document}